\documentclass[twocolumn,prl,showpacs,preprintnumbers,amsmath,amssymb]{revtex4}
\usepackage[dvipdfm]{graphicx}
\usepackage{dcolumn}
\usepackage{bm}
\usepackage{amssymb}

\begin{document}
\title{Supersolid phase in spin dimer XXZ systems under magnetic field}

\author{Kwai-Kong Ng}
\author{T. K. Lee}
\affiliation{Institute of Physics, Academia Sinica, Nankang,
Taipei 11529, Taiwan}
\date{\today}
\begin{abstract}
Using quantum Monte Carlo method, we study, under external magnetic
fields, the ground state phase diagram of the two-dimensional spin
$S$=1/2 dimer model with an anisotropic intra-plane
antiferromagnetic coupling. With the anisotropy $4 \gtrsim \Delta
\gtrsim 3$, a supersolid phase  characterized by a non-uniform bose
condensate density that breaks translational symmetry is found. The rich
phase diagram also contains a checkerboard solid and two different
types of superfluid phase formed by $S_z=+1$ and $S_z=0$ spin
triplets, with finite staggered magnetization in z-axis and in-plane
direction, respectively. As we show, the model can be realized as a
consequence of including the next nearest neighbor coupling among
dimers and our results suggest that spin dimer systems may be an
ideal model system to study the supersolid phase.
\end{abstract}


\maketitle
Five decades ago, Penrose and Onsager \cite{Penrose} showed that the
coexistence of both off-diagonal long-range order(ODLRO) and
diagonal long range order (DLRO), i.e. a supersolid (SS) state, is
impossible in commensurate solids. However, it was later
demonstrated \cite{Andreev} that vacancies in solid may still exhibit
Bose-Einstein condensation. And more recently, Prokof'ev and Svistunov
\cite{Prokofev} provide a proof that the presence of vacancies
and/or interstitial atoms is a necessary condition for a SS. They
therefore exclude the possibility of a bulk SS in the recent
experimental observations of nonclassical moment of inertia in solid
$^4$He \cite{Kim}. Instead, interpretation based on superfluid
interfaces is suggested. While experimental evidence of SS is still
inconclusive, numerical stimulations have already confirmed the
presence of SS of bosons in various conditions, that is including
next nearest neighbor interactions (n.n.n.) \cite{Batrouni} or with
softcore \cite{Sengupta} in square lattices, and with hardcore in
triangular lattices \cite{Wessel1}.

All of the models discussed above concentrate on studying bosons.
Knowing the possibility of representing bosons by spins or vice
versa, we wonder if SS phase also exists in spin systems. Recent
experiments on spin dimer compounds, such as TlCuCl$_3$, KCuCl$_6$
and BaSiCu$_2$O$_6$ \cite{Ruegg}, have confirmed the formation of super fluid (SF)
of the spin triplets under an external magnetic field. These spin
bosons have also previously been found to be crystallized, signaled
by the magnetization plateau, in dimer compounds such as
SrCu$_2$(BO$_3$)$_2$\cite{Kageyama}. The competition between
repulsive interaction and kinetic motion decides whether the ground
state is crystalized or is a SF \cite{Rice}. This also raises the
possibility of a SS state. While SS is unstable against phase
separation for hardcore boson in a square lattice
\cite{Sengupta,Schmid}, the spin bosons are shown to be
semi-hardcore \cite{Ng}. In this work, using quantum Monte Carlo
(QMC) method we demonstrate that the ground state of a spin dimer
XXZ model could be SF of two kinds of triplets, a quantum solid or,
most importantly, a SS. This XXZ anisotropy can be easily realized
if the n.n.n. interaction among dimers is considered. Therefore,
while previous proposal of detecting SS in optical lattices
\cite{Wessel1} is technically difficult, we suggest that, due to the
semi-hardcore nature of spin bosons, it may be more natural to look
for a SS in spin dimer systems. Furthermore, we provide strong
evidence that in the SS phase, the SF condensate amplitude has a
spatial modulation and possesses a lattice symmetry on its own. This
kind of non-uniform condensate has been discussed related to solid
$^4$He \cite{Kumar} but has not been confirmed in numerical studies
before.

In particular, we study the 2D bilayer AF spin dimer XXZ model
\begin{eqnarray}
H_{XXZ}&=&- h \sum_{\alpha,i} S^z_{\alpha,i}+J \sum_i {\bf S}_{1,i}
\cdot
{\bf S}_{2,i} \\
&&+ J' \sum_{\alpha,\langle i,j \rangle}
(S^x_{\alpha,i}S^x_{\alpha,j}+S^y_{\alpha,i}S^y_{\alpha,j}+\Delta
S^z_{\alpha,i}S^z_{\alpha,j}),\nonumber
\end{eqnarray}
where $\alpha=1,2$ denotes the layer index and the third term sums
over all nearest neighbors (n.n.) in the $xy$ plane for both layers.
Since $J>>J'$ (we take $J'/J=0.29$ in this paper), the low energy
physics is conveniently described by the interaction of the four
spin states, singlet ($|s\rangle$) and triplets $S^z=0$, $\pm 1$
($|t_{0,\pm}\rangle$) of each dimer.

When $\Delta=1$, i.e., the isotropic case, the model successfully
describes the SF phase, characterized by the in-plane staggered
magnetization $m_{xy}$, of the spin 1/2 dimer compounds TlCuCl$_3$,
KCuCl$_6$, and BaSiCu$_2$O$_6$ \cite{Matsumoto,Ng}. The mean field
(MF) ground state is
\begin{eqnarray}
|\Psi_0\rangle &=& \prod_{i} \left[ u |s\rangle_i + (-1)^i(v e^{i
\theta}|t_+\rangle_i + w e^{-i\theta}|t_-\rangle_i) \right],
\label{wf}
\end{eqnarray}
where $u^2+v^2+w^2=1$ and we have neglected $|t_0\rangle$ states
here. The order parameter is given by $m_{xy}=u(v+w)$ whose square
is shown to be equal to the condensate density ($n_0$, the zero
momentum number density) of the semi-hardcore boson $b^\dag_i =
\frac{(-1)^i}{\sqrt{2}} \left( S^+_{1,i} - S^+_{2,i} \right)$, which
allows up to double occupancies on each site. The global phase
$\theta$ corresponds to the orientation of the $m_{xy}$ in the $xy$
plane. Ignoring $|t_-\rangle$, i.e. let $w=0$, it is a hardcore
model with n.n. repulsive interaction $V$ and hopping $t$. For the
case of square lattice, $V=t=J/2$. This Hamiltonian preserves the
particle-hole symmetry, which is broken when $|t_-\rangle$ is taken
into account and thus an asymmetric phase diagram of $m_{xy}(h)$ is
observed.

By increasing $\Delta>1$, the repulsive interaction between the
triplets enhances. The ratio of interaction energy and kinetic
energy is determined by $\Delta=V/t$. Therefore, a large value of
$\Delta$ will lead to a solid phase where the structure factor has a
finite peak at certain wave vector. The compound
SrCu$_2$(BO$_3$)$_2$ \cite{Kageyama} has shown such solid phase in
its magnetization plateaus. When both interaction energy and kinetic
energy play crucial roles, the coexistence of both SF ordering and
solid ordering is possible. Our numerical simulation indeed shows
the presence of SS phase from $\Delta\approx 3$ to 4. In our QMC
simulation, we employ the stochastic series expansion approach
\cite{Sandvik} on a 2D bilayer square lattice for different linear
sizes $L=10$, 12, 16 and 20, with temperature $T$ inversely
proportional to $L$ such that at $L=20$, $T/J=0.02154$.

\begin{figure}
\includegraphics[width=7cm]{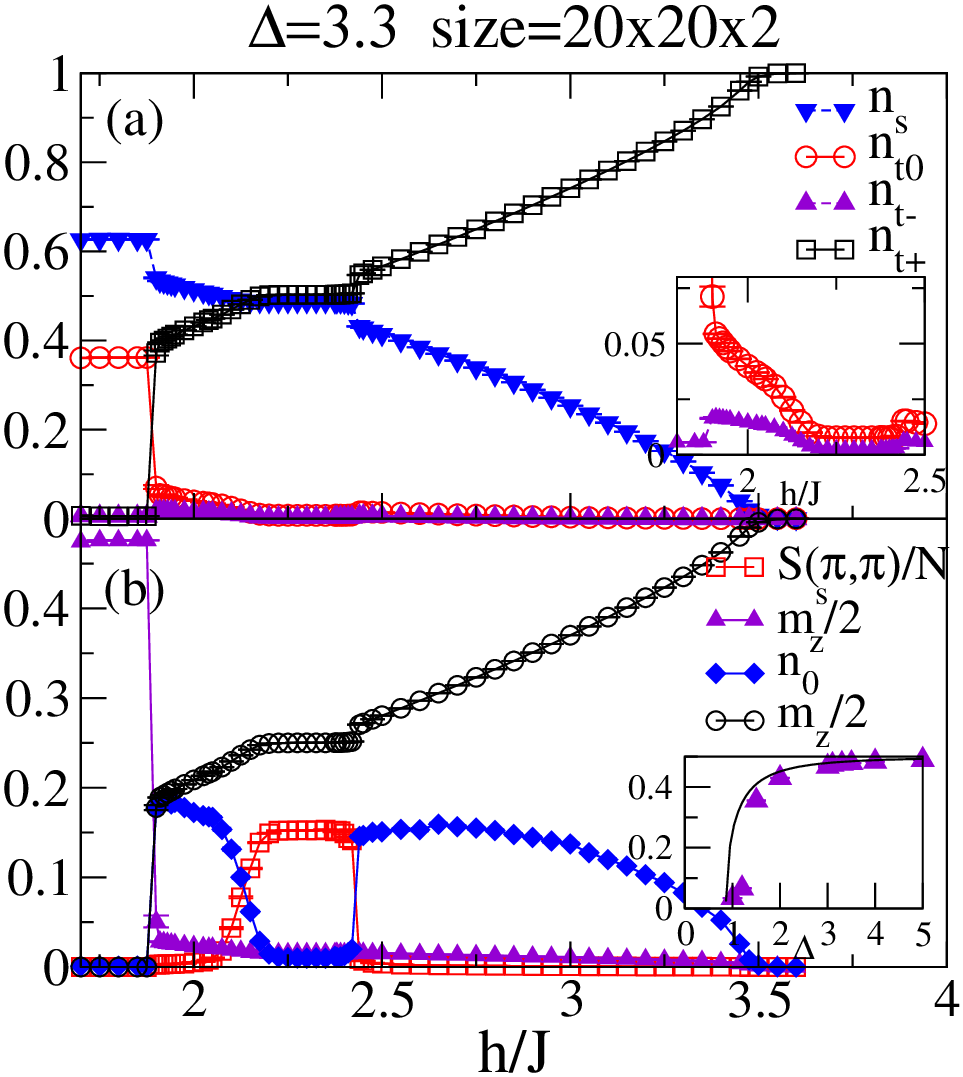}
\caption{(color online) Phase diagrams of (a) the number densities
$n_s$, $n_{t+}$, $n_{t0}$, and $n_{t-}$, and (b) the order
parameters $n_0$, $S(\pi,\pi)/N$, $m_z/2$, and $m^s_z/2$. The upper
inset shows the details of $n_{t-}$ and $n_{t0}$ around the SS phase
with enlarged scale. The lower inset compares the QMC (symbols) and
the MF (solid lines) results of $m^s_z/2$ as a function of $\Delta$.
} \label{fig1}
\end{figure}

In Fig.\ref{fig1}, we show the number densities (1a) and order parameters (1b) as a function of
magnetic field $h$ for $\Delta=3.3$. Let us discuss the different
phases starting from low field $h$.

\emph{superfluid I} (SFI) -- At low field, the phase is
characterized by the only non-vanishing order parameter, the
staggered magnetization along z-axis $m^s_z=\langle \sum_i (-1)^i
(S^z_{1i}-S^z_{2i}) \rangle/N$ (N is the number of dimers), which is
field independent. This is a SF phase forming by $|t_0\rangle$ state
in the background of $|s\rangle$ state with a MF wavefunction
$|\Psi_{t0}\rangle=\sum_i (u_0 |s\rangle_i + e^{i \phi}(-1)^i v_0
|t_0\rangle_i)$, where $u_0$ and $v_0$ are given by minimizing the
MF energy and have the form $u_0^2 = \frac{1}{2} (1+\frac{J}{4
\Delta J'})$ and $v_0^2 = \frac{1}{2} (1-\frac{J}{4 \Delta J'})$.
Then, $m^s_z=2 u_0 v_0 \cos \phi=  \langle b_0^\dag \rangle +
\langle b_0 \rangle$, with $b^\dag_0$ is the hardcore boson operator
that creates a $|t_0\rangle$ state. At finite field, the global
phase $\phi$ is fixed to zero so that the largest $m^s_z$ aligned to
the field direction. The inset of Fig.\ref{fig1}b shows an excellent
agreement of MF $m^s_z/2$ to the values obtained from QMC for
different $\Delta$'s. Even for $\Delta=1$, that is relevant to many
spin dimer materials, as long as $1 > J'/J> 1/4$, so $v_0 > 0$,
there is still a small but finite $m^s_z$ in this low field phase.

\emph{Superfluid II} (SFII) -- When $h/J \gtrsim 1.87$, the strong
Zeeman energy favors $|t_+\rangle$ instead of $|t_0\rangle$ and
leads to the condensation of $|t_+\rangle$. In Fig.\ref{fig1}(b),
 the condensate density $n_0=m^2_{xy}$ of boson $b^+$, jumps
abruptly to about 0.2 as $m^s_z$ drops to a small value. Note that
$n_{t0}$ is small but still finite so that $|t_0\rangle$ behaves as
impurities. This phase reappears at even higher fields.

\begin{figure}
\includegraphics[width=6.5cm]{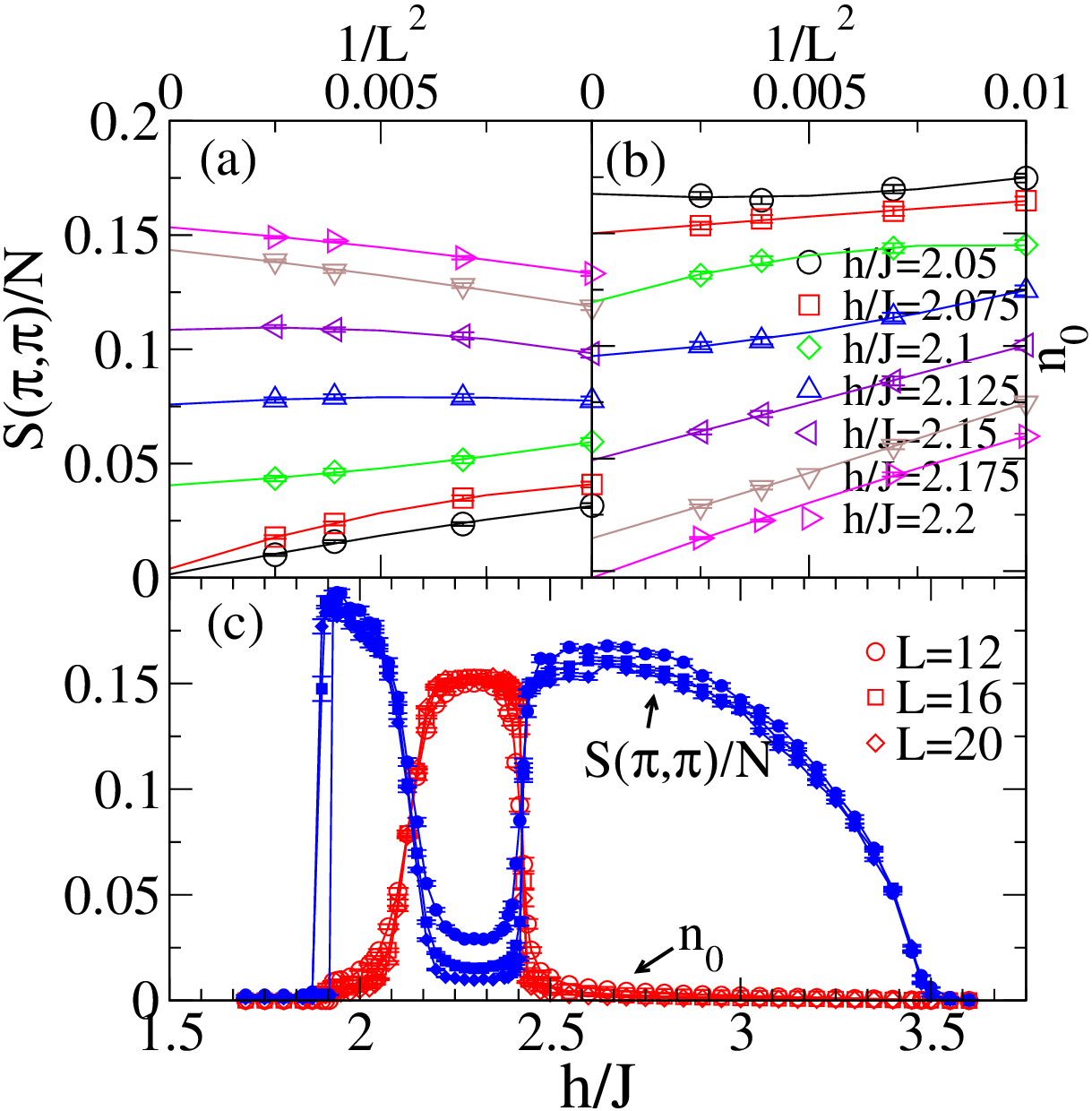}
\caption{(color online) Finite size scaling of (a) $S(\pi,\pi)/N$
and (b) $n_0$. In (c), results of $S(\pi,\pi)/N$ (open symbols) and
 $n_0$ (filled symbols) for $L=12$, 16 and 20.}\label{fig2}
\end{figure}
\begin{figure}
\includegraphics[width=6cm]{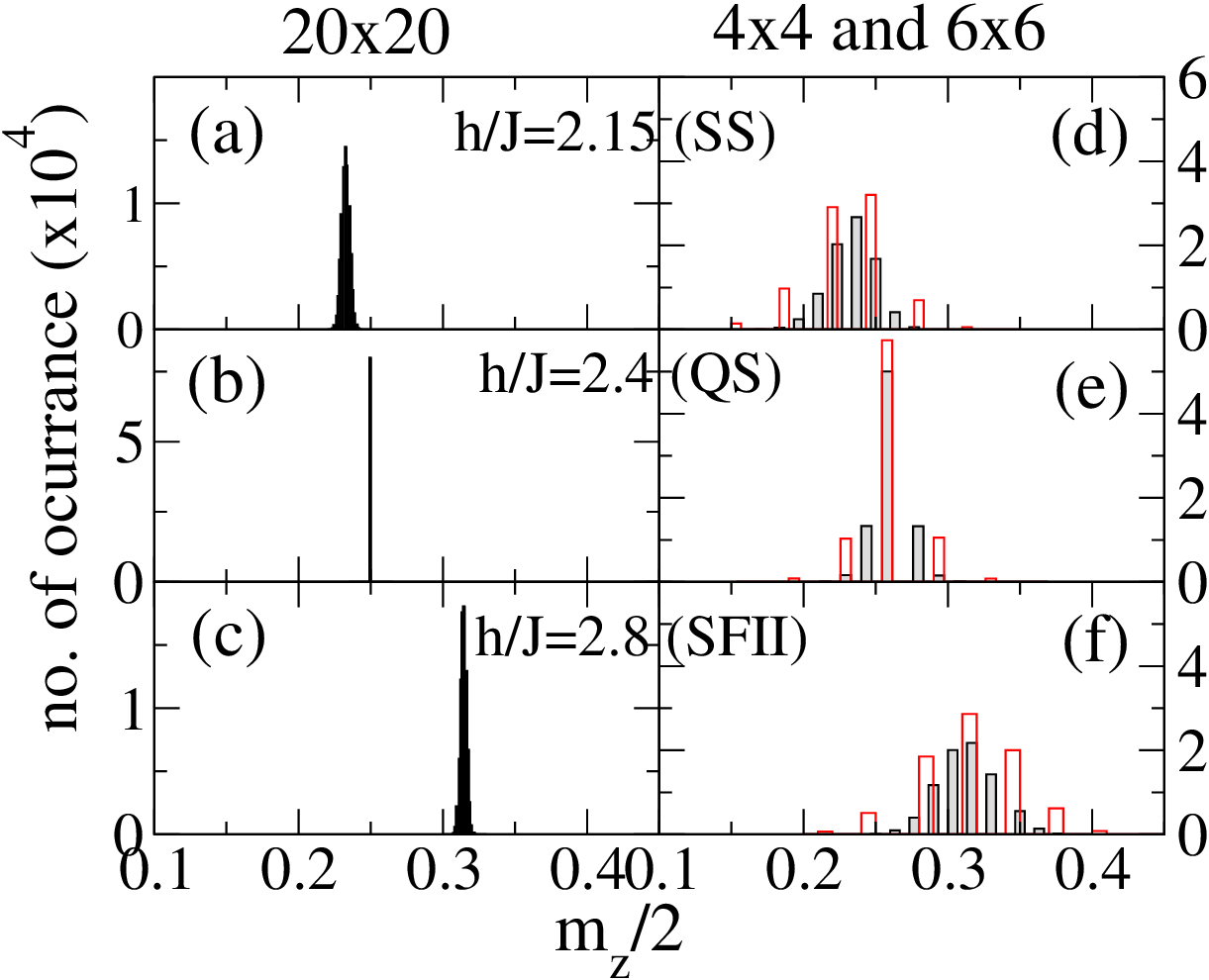}
\caption{(color online) Histograms of the averaged $m_z/2$ of the
whole lattice ((a) to (c)) and of a small area ((d) to (f)) of size
4x4 (unshaded) and 6x6 (shaded). $\Delta=3.5$. For comparison,
histograms of SFII and QS are also presented.} \label{fig3}
\end{figure}

\emph{Supersolid} (SS) -- At $h/J\sim 2.10$, $n_0$ begins to reduce
rapidly while the structure factor $S(\textbf{Q})/N=\sum_{ij}
\langle n_{i,t+} n_{j,t+} e^{i
\textbf{Q}\textbf{r}_{ij}}\rangle/N^2$ at the wave vector
$(\pi,\pi)$ of $|t_+\rangle$ increases. The single peak of
$S(\pi,\pi)/N$ implies a checkerboard solid ordering of
$|t_+\rangle$ in the system. The transition from SFII to SS is also
signaled by the kinks observed in the $m_z$, $n_0$, $n_{t-}$, and
$n_{t0}$ (Fig.\ref{fig1}). Within $2.2 \gtrsim h/J \gtrsim 2.1$,
both SF and solid order parameters remain finite and constitute the
SS phase. Finite size scaling shown in Fig.\ref{fig2} indicates that
the SS phases is stable in the thermodynamic limit. To address the
concerns of phase separation, i.e. the possibility of having
separate domains of checkerboard solid and SF, we measure the
average $m_z$ within a small area of size 6x6 and 4x4 of a 20x20
lattice. Since $m_z$ differs for different domains, the small area
measurement should give two distinct peaks in the histogram in the
case of phase separation. Shown in Fig.\ref{fig3} are the histograms
of $m_z/2$ for selected fields. The distribution is wider in 6x6
case when compared to that of whole 20x20 lattice where fluctuation
is averaged out. The single Gaussian peak at $h/J=2.15$ indicates
that there is no domains of different magnetization and rules out
the possibility of phase separation.

\emph{Quantum solid} (QS) -- Increasing the field further
($h/J>2.20$) will half-fill the system with $b^+$ so that $m_z=0.5$.
Because of strong repulsive interaction between nearest neighbor
$b^+$, adding or removing extra boson to this checkerboard solid
state requires a finite energy. This causes the stability of the
state and the plateau in the $m_z$ curve. Differing from the
classical value $S(\pi,\pi,)/N = 1/4$, the calculated structure
factor in the solid phase is about 0.13 for $\Delta=3.3$. The $m_z$
of each sublattice are found to be 0.890 and 0.110 respectively
(Table \ref{table1}).  As $\Delta$ increases, $S(\pi,\pi)/N$
approaches the classical value of 1/4. At higher fields, the melting
of the solid is of first order and the SFII phase reappears.
$n_{t+}$ keeps increasing while all other states reduces to zero at
$h=3.50$, where all the spins are fully polarized.

\begin{figure}
\includegraphics[width=5cm]{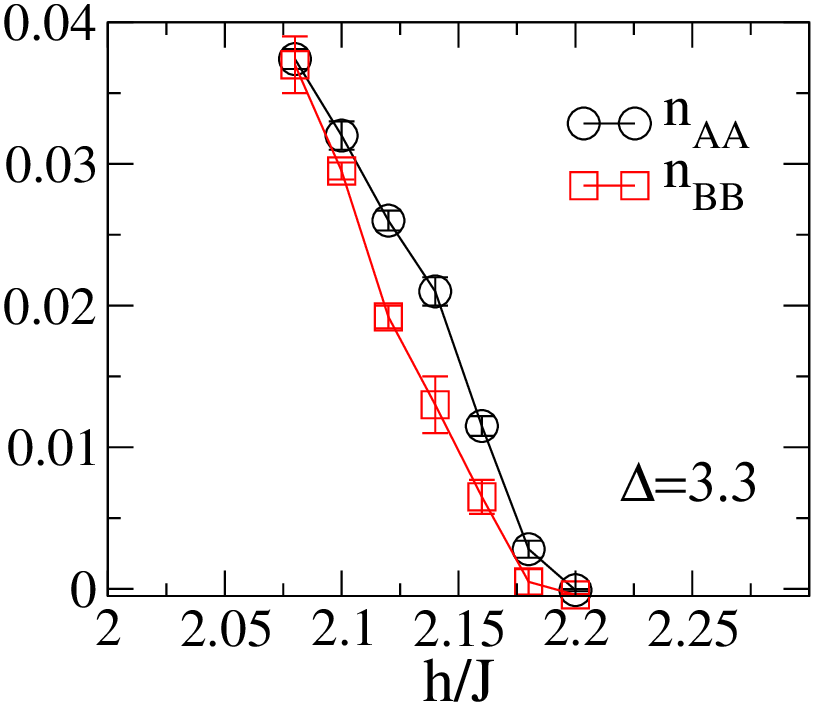}
\caption{(color online) Sublattice condensates $n_{AA}$ and $n_{BB}$
in the SS phase extrapolated to $T\rightarrow 0$ and
$L\rightarrow\infty.$ }\label{fig4}
\end{figure}

Here we address the question of non-uniform condensate in the SS
phase. Table \ref{table1} shows that the $m_z$ differs in different
sublattices A and B, and breaks the translation symmetry. To examine
the condensate more closely, we define the sublattice condensates
$n_A$ and $n_B$ as $n_\alpha= \sum_{i \in \alpha} \langle b_i^\dag
\rangle/N$, where $\alpha=A,B$. In our simulation, we compute the
products of sublattice condensate $n_{\alpha \beta}=\sum_{i \in
\alpha, j \in \beta} \langle b_i^\dag b_j \rangle/N^2$. As shown in
Table \ref{table1}, $n_{AA}\not=n_{BB}$ and
$n_{AB}=n_{BA}=\sqrt{n_{AA}n_{BB}}$ in the SS phase. Thus $n_{\alpha
\beta}=n_\alpha n_\beta$, and $n_A=0.130$, $n_B=0.148$. It indicates
that the condensate has a checkerboard ordering on its own, which is
the intrinsic nature of a SS state \cite{Kumar}. This is different
from the idea of having a SF component "on top" of a bose solid. The
checkerboard order of the condensate disappears in the usual
superfluid phase SFII, while the condensate amplitude on both
sublattices vanishes in the QS phase as expected. The result for
different $h$ is plotted in Fig. \ref{fig4}. The SS phase, instead
of being a crossover between SFII and QS, is clearly a new phase
with second order phase transitions from SFII and QS respectively.
Nevertheless, a more detail study of scaling behavior is needed to
determine the scaling exponents. This result also rules out phase
separation.
\begin{table}[!b]
\begin{tabular}{c|c|c|c|c|c|c}
& $m_z^A/2$ & $m_z^B/2$ & $n_{AA}$ & $n_{BB}$ & $n_{AB}$ & $n_{BA}$ \\
\hline
 SFII & 0.224(5) & 0.225(5) & 0.0384(7) & 0.0384(6) &
0.0384(7) &
0.0387(6) \\
SS & 0.397(1) & 0.0831(9) & 0.0169(5) & 0.0219(5) & 0.0195(5) &
0.0194(4) \\
QS & 0.4449(2) & 0.0549(2) & 0.0017(1)& 0.0036(1)& 0.0026(1) &
0.0027(1)
\end{tabular}
\caption{Sublattice magnetizations and condensates for SFII
($h/J$=2.08), SS ($h/J=$2.14) and QS ($h/J=2.24$) phases.
$\Delta=3.3$ and L=20. The small values of $n_{\alpha \beta}$ in QS
phase is due to the finite size effect.} \label{table1}
\end{table}
\begin{figure}
\includegraphics[width=6.5cm]{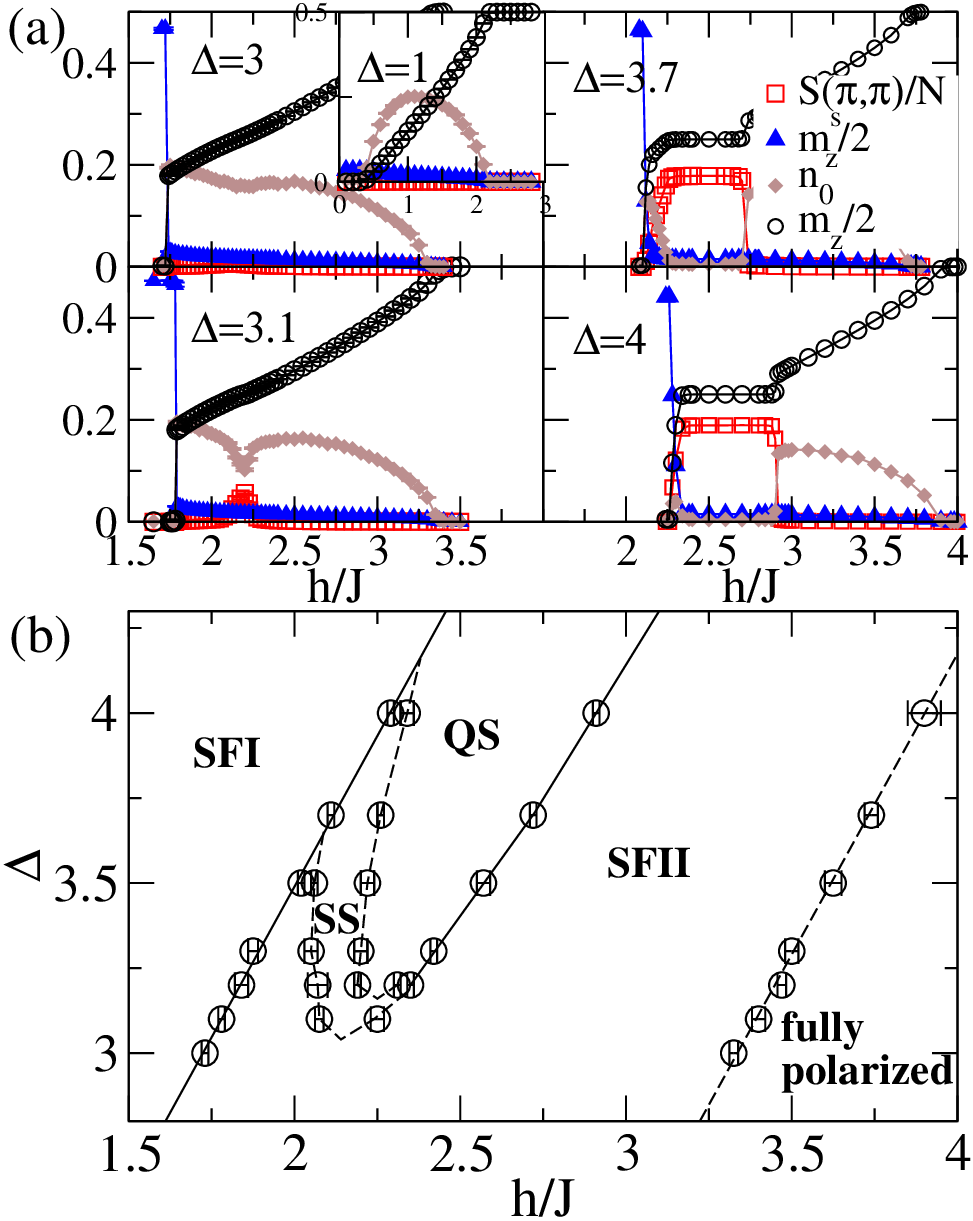}
\caption{(color online) (a) Order parameters as a function of $h/J$
for different $\Delta$ with lattice size 20x20x2. The inset shows
the isotropic case ($\Delta=1$). (b) Ground state phase diagram
(extrapolated to $L\rightarrow\infty$) of $\Delta$ vs. $h/J$. Solid
(dotted) lines denote the first (second) order phase boundaries.}
\label{fig5}
\end{figure}

Fig.\ref{fig5}(a) and (b) show the plots of order parameters as a
function of $h$ for different $\Delta$ and the phase diagram of
$\Delta$ vs. $h$, respectively. The isotropic case, $\Delta=1$, is
similar to those observed experimentally on other spin dimer
compounds. Enhancing the anisotropy to $\Delta=3$, there is a small
drop in $n_0$ around $m_z=0.5$ while $S(\pi,\pi)/N$ remains zero.
For even larger repulsive interaction at $\Delta=3.1$, the drop
develops further with a finite peak of $S(\pi,\pi)/N$ at $m_z=0.5$.
Note that it is a commensurate SS that is made possible by the
semi-hard core nature of the spin bosons. The finite peak of
$S(\pi,\pi)/N$ finally develops to a plateau together with the $m_z$
in the QS phase. As a consequence of lowering energy of
$|t_0\rangle$ state with increasing $\Delta$, the phase boundary of
SFI to SFII moves to the higher fields and shrinks the region of
SFII and SS which eventually vanishes at $\Delta \sim 4$ and leaves
a first order transition from SFI to QS phase. As shown in
Fig.\ref{fig5}(b), the SS phase is stable in a parameter region with
$4 \gtrsim \Delta \gtrsim 3$.

The next question is whether these phases can be realized in real
material. There are two issues, however, that have to be considered.
The first is the anisotropy that introduced in $H_{XXZ}$ only
applies to the interdimer coupling but not the intradimer coupling.
This kind of selective anisotropy is hard to realize. The second
issue concerns the large value of $\Delta \sim 3$ for the SS to be
stable. In usual compounds the XXZ anisotropy is caused by the
spin-orbit coupling, which couples the spin to the crystal
structure, and the effective $\Delta$ is much smaller than 3. To the
first issue, one may instead consider a $S=1$ compound with a single
ion anisotropy $D\sum (S^z)^2$, where $D$ plays the role of
intra-dimer coupling $J$ and the states $S^z=\pm 1$ and $S^z=0$
plays the role of $|t_{\pm}\rangle$, and $|s\rangle$, respectively.
There is no corresponding states of $|t_0\rangle$. Therefore, SFI is
replaced by $S^z=0$ state on each site. Together with spin-orbit
coupling that realizes the XXZ anisotropy, one has essentially the
same phase diagram as in the case of $H_{XXZ}$, although a large
enough anisotropy is still needed. On the other hand, there is
another route to look for the realization of the SS phase. Consider
two neighboring $t_+$ states, the inclusion of n.n.n. coupling
$J''{\bf S}_{1,i} {\bf S}_{2,j}$ ($i,j$ are n.n.) will simply
enhance the repulsion between $|t_+\rangle$ states. Contrastingly,
if $|t_0\rangle$ is ignored, $J''$ reduces the hopping between the
neighboring $|t_+\rangle$ and $|s\rangle$ states due to the
antisymmetry of $|s\rangle$. In bond operator representation, when
the $|t_0\rangle$ states is ignored, spin operators
$S^z_{1,2}\approx\frac{1}{2}(t^\dag_+ t_+ - t^\dag_- t_-)$ and
$S^+_{1,2}\approx\pm \frac{1}{\sqrt{2}} (s^\dag t_- -t^\dag_+ s)$.
Then the n.n.n. coupling is approximately given by
\begin{equation}
{\bf S}_{1,i} {\bf S}_{2,j} \approx S^z_{1,i} S^z_{1,j}-\frac{1}{2}
(S^+_{1,i} S^-_{1,j} + S^-_{1,i} S^+_{1,j}).
\end{equation}
The inclusion of $J''$ now breaks the spin rotational symmetry and
leads to an XXZ model with effective n.n. coupling of $J^*=J'-J''$
and anisotropy $\Delta$ given by $J^* \Delta=J'+J''$. Therefore, if
$J''=J'/2$ one has a spin dimer XXZ model with $\Delta=3$, close to
the SS phase.

In summary, we demonstrate that the spin dimer XXZ model, a natural
semi-hardcore boson system with defects, contains a SS phase that
characterized by the biparte condensate density. The anisotropy can
be a consequence of including the n.n.n. coupling among dimers.
While it is suggested \cite{Wessel1} that SS state of hardcore
bosons can be tested in triangular optical lattices, we propose that
spin dimer compounds may be a natural place to realize the SS state
of semi-hardcore bosons.
\begin{acknowledgments}
 We acknowledges financial support by the NSC
(R.O.C.), grant no. NSC 94-2112-M-001-003.
\end{acknowledgments}

\end{document}